\documentclass[10pt,twocolumn,showpacs,longbibliography,amsmath,amssymb,osajnl,floatfix,superscriptaddress]{revtex4-2}

\usepackage{graphicx}
\usepackage{dcolumn}
\usepackage{bm}
\usepackage{color}
\usepackage{txfonts}
\usepackage{microtype}
\usepackage{mathrsfs}
\usepackage{bm}
\usepackage{setspace}
\newcommand\numberthis{\addtocounter{equation}{1}\tag{\theequation}}

\begin{document}
\author{Yue-Yue Chen}\email{yueyuechen@shnu.edu.cn}
\affiliation{Department of Physics, Shanghai Normal University, Shanghai 200234, China}		
\author{Karen Z. Hatsagortsyan}
\affiliation{Max-Planck-Institut f\"{u}r Kernphysik, Saupfercheckweg 1, 69117 Heidelberg, Germany}
\author{Christoph H. Keitel}\affiliation{Max-Planck-Institut f\"{u}r Kernphysik, Saupfercheckweg 1, 69117 Heidelberg, Germany}
\author{Rashid Shaisultanov}\email{r.shaisultanov@hzdr.de}
\affiliation{Helmholtz-Zentrum Dresden-Rossendorf, Bautzner Landstraße 400, 01328 Dresden, Germany}

\title{Electron spin- and photon polarization-resolved probabilities of strong-field QED processes}

\date{\today}

\begin{abstract}

A derivation of fully polarization-resolved probabilities is provided for high-energy photon emission
and electron-positron pair production 
 in ultrastrong laser fields. The probabilities resolved in both electron spin and photon polarization of incoming and outgoing particles are indispensable for developing QED Monte Carlo and QED-Particle-in-Cell codes, aimed at the investigation of polarization effects in nonlinear QED processes in ultraintense laser-plasma and laser-electron beam interactions, and other nonlinear QED processes in external ultrastrong fields, which involve multiple elementary processes of a photon emission and pair production. The quantum operator method introduced by Baier and Katkov is employed for the calculation of probabilities within the quasiclassical approach and the local constant field approximation. The probabilities for the ultrarelativistic regime are given in a compact form and are suitable to describe polarization effects  in strong laser fields of arbitrary configuration, rendering them very well suited for  applications.

\end{abstract}

\maketitle

\section{Introduction}

The investigation of spin dynamics of leptons driven by external fields and the polarization characteristics of their emissions have important implications in many fields, including high-energy 
\cite{anthony2004observation,moortgat2008polarized} and nuclear physics \cite{horikawa2014neutron,uggerhoj2005interaction,abe1995precision,alexakhin2007deuteron}, and material science \cite{kessler2013polarized,getzlaff2010experimental}. Apart from the potential of generating polarized ultrarelativistic particle beams for various applications, for instance, spin polarized electron (positron) beams for probing nuclear
structure and new physics beyond the standard model \cite{borisov1996compton,herczeg2003cp,ananthanarayan2018inclusive,godbole2006lepton},
or $\gamma$-photon beams
for  meson photoproduction \cite{akbar2017measurement} and vacuum birefringence measurement in ultrastrong laser fields \cite{nakamiya2017probing,bragin2017high,king2016vacuum,ilderton2016prospects,ataman2017experiments}, the understanding of the polarization dependence of nonlinear Compton scattering  and Breit-Wheeler processes is of great interest for modelling high-order QED effects such as the trident process \cite{hu2010complete,ilderton2011trident,king2013trident} and double nonlinear Compton scattering \cite{morozov1975elastic,seipt2012two,mackenroth2013nonlinear,king2015double}, as well as polarized QED cascades \cite{seipt2021polarized,nerush2011laser}.

The  problem of radiation by an ultrarelativistic electron in a strong laser field can be split into two characteristic regimes depending on the classical strong-field parameter $a_0=eE_0/(m\omega_0)$ \cite{Ritus1985,Baier1998}. For $a_0\lesssim 1$,  the total angle of the electron deflection in the external field ($\sim a_0/\gamma$)  is lower or of the order of the characteristic angle of  radiation ($\sim 1/\gamma$),
and the radiation of the particle is determined by a significant part or nearly the whole trajectory of the particle.
In this case, the characteristics of radiation is more sensitive to the features of the external field, such as the pulse shape and polarization \cite{seipt2011nonlinear,heinzl2010beam,bocquet1997graal}. Here, $E_0$ and $\omega_0$ are the laser field and frequency, respectively, $-e$, $m$, and $\gamma$ the electron charge, mass, and the Lorentz-factor, respectively, while relativistic units with $c=\hbar=1$ are used throughout.

Recent progress in laser technology \cite{Yoon_2021,Burdonov_2021,Garcia_2021,Hong_2021,Vulcan,ELI,XCELS} enables observation of nonlinear QED processes in ultrastrong fields and has stimulated the interest of theoretical investigations
to the highly nonperturbative domain with $a_0\gg 1$ \cite{Piazza2012,Heinzl_2012}.
The radiation spectra in the strong field regime can be calculated in the Furry picture within the quantum theory if the solution of wave equations in the given external field is known, see e.g. \cite{seipt2020spin,king2020nonlinear,wistisen2014interference,mackenroth2011nonlinear,Dinu_2020,Torgrimsson_2021}.
 However, such solutions are known only for a few specific fields \cite{Bagrov_1990} and not in most realistic field configurations.
The Volkov wave function for a relativistic  electron in a monochromatic plane wave field \cite{Volkov_1935} has been fully exploited within the Furry picture for calculations as polarization averaged \cite{Ritus1985}, as well as for polarization-resolved processes. In particular, the  electron spin-resolved radiation probability is calculated in Refs.~\cite{ritus1972radiative,bol2000spin} with averaging over the emitted photon polarization, and the photon polarization resolved probability in Refs.~\cite{king2013photon,king2020nonlinear,tang2020highly}, averaging over the   electron spin variable. Orbital angular momentum transfer in the nonlinear Compton process is discussed in \cite{chen2018gamma}.
A comprehensive description of polarization dependent nonlinear Compton scattering in
a monochromatic plane-wave background has been given in Ref.~\cite{ivanov2004complete}, including both the electron spin and photon polarization, which however yields rather unwieldy analytical expressions for probabilities as a sum over high-order Bessel functions and are  difficult to apply in QED-PIC codes. Recently, polarization resolved probabilities in plane-wave laser pulses have been numerical evaluated in \cite{seipt2020spin}.

In ultrastrong field regime  one can employ the approximate asymptotic expressions for probabilities at $a_0\gg 1$. In physical terms this approximation stems from the fact that the  formation length of the process becomes much smaller in this limit than the typical scale of the trajectory: $l_f\sim\lambda_0/a_0\lesssim \lambda_0$ \cite{Ritus1985}. In other words, the  total angle of the particle deflection in external fields  is much larger than the characteristic angle of  radiation, and in the given direction
the particle radiates from a small fraction of its trajectory.
In this case, the variation of the external field acting on the particle within the formation length can be neglected, leading to the local constant field approximation (LCFA) \cite{Ritus1985,di2012extremely,seipt2020spin,seipt2018theory,king2020nonlinear,di2018implementing,di2019improved,lv2021anomalous}.  More accurate conditions for the LCFA are $a_0\gg1$ and $a_0^3/\chi_e\gg1$, which stem  from the saddle-point approximation in calculations of the time-integral for the amplitude of the process \cite{dinu2016quantum,ilderton2019extended}.
Here, $\chi_e=|F_{\mu\nu}p^\nu|/mF_{cr}$ with $F$ being the field strength tensor, $p$  electron momentum, and $F_{cr}=m^2/|e|=1.3\times10^{16}$~V/cm the critical field of QED.

The collisions of a strong laser field and high-energy particles also enable the production of $e^-e^+$ pairs, which has been successfully observed at the Stanford Linear Accelerator Center (SLAC) in 1990s \cite{bamber1999studies,burke1997positron}. The creation of pairs has been attributed to the nonlinear Breit-Wheeler process, where the single $\gamma$-photon absorption is accompanied with a simultaneous absorption of multiple laser photons. The spin effects in this process in a monochromatic plane laser wave have been analyzed in Refs.~\cite{villalba2013photo,jansen2016strong}, averaging over the $\gamma$-photon polarization,
while the photon polarization effects have been studied in Refs.~\cite{nikishov1964zh,nikishov1967pair,ritus1972vacuum}, averaging over the electron-positron spins.
The same processes in a constant crossed field have been considered in Refs.~\cite{nikishov1967pair,ritus1970radiative,king2013photon}.
A more comprehensive analytical treatment of the nonlinear Breit–Wheeler process in a monochromatic  plane-wave laser field, including both electron-positron spins and photon polarizations, has been presented in Ref.~\cite{ivanov2005complete} (the description of this process  via  helicity amplitudes is given in \cite{tsai1993laser}), and the numerical analysis of the process in Ref.~\cite{seipt2020spin}.

The semiclassical QED operator method has been developed by Baier and Katkov \cite{Baier1998} for efficient calculations of probabilities of strong-field QED processes in strong background fields, and provides a powerful major alternative to the QED calculations in the Furry picture. The QED operator method is applicable when the electron dynamics in the external field is quasiclassical (amenable to the  Wentzel-Kramers-Brillouin approximation), however it accounts fully for the quantum features of the QED process, in particular, the  photon recoil at radiation, as well as the possibility of the pair creation by a $\gamma$-photon. The amplitude of the QED process in the operator method is derived assuming commutativity  of operators describing the particles due to the quasiclassical dynamics, and taking into account the noncommutativity of the particle operators with those of the photon field \cite{Landau_4}. Finally, the process amplitude is derived as a functional of the electron classical trajectory in the given background field. Especially simple analytical expressions for the amplitude are obtained in the ultrarelativistic regime, applying the $1/\gamma$-expansion up to the leading order, which provides the process description within the LCFA. Recently, the semiclassical QED operator method beyond LCFA has been applied numerically to investigate the polarization effects in laser fields of moderate intensity \cite{wistisen2020numerical}, where the calculation of radiation spectra was carried out with  numerical integrations using the electron exact classical  trajectories.

In this paper, we derive the spin- and polarization-resolved radiation and pair production probabilities applicable for investigations of polarization effects in realistic ultrastrong laser fields with $a_0\gg 1$. The fully  polarization-resolved quasiclassical formulas are obtained using the QED
operator method of Baier and Katkov within the LCFA. While in the seminal book
by Baier, Katkov, and Strakhovenko \cite{Baier1998},  the radiation  and pair production probabilities are given only for the case when the spin state of one of the outgoing particles is summed over,
here we obtain the fully polarization resolved formulas and without specification of the spin quantization axis.
These probabilities are indispensable to develop Monte Carlo codes applied for detailed investigations of polarization phenomena in QED processes in ultrastrong laser fields \cite{Liyf2019,Chen2019,Wan2020,li2020production,Liyf2020,Wanf2020}, in particular, during nonlinear Compton scattering and nonlinear Breit-Wheeler processes. In our previous publications \cite{Liyf2019,Chen2019,Wan2020}, we have used a spin-resolved but photon polarization averaged QED Monte Carlo code. In \cite{li2020production} the Monte Carlo code was based on the probabilities averaged over the outgoing particle polarization, while in Refs.~\cite{Liyf2020,Wanf2020} we used the fully polarization-resolved probabilities, however without giving the derivation of corresponding formulas. The aim of this paper is to provide the derivation of the fully polarization-resolved probabilities, allowing their straightforward verification and a reliability check.

\section{Spin and polarization resolved radiation probability}

The problem of radiation of ultrarelativistic electrons  in an external electromagnetic field can be solved with the quasiclassical operator approach, developed by Baier and Katkov \cite{BKp67,BK67} and inspired by \cite{Schwinger_1954}. It is based on the analysis of two types of quantum effects at the radiation of high-energy particle
in an external field. The first type originates from the quantization of particle
motion in the field. The latter yields noncommutativity  of operators of the particle dynamical variables, with the nonvanishing order of the commutator scaling as $\chi/\gamma^{3}$ (for instance in a constant magnetic field).
Therefore, at high energies $\gamma\gg 1$ ($\chi\lesssim 1$) the motion of the particle is quasiclassical.
The second type of quantum effects is related to the quantum recoil of a particle (with an energy $\epsilon$) during a photon emission (with an energy $\omega$) and it is of the order $ \omega/\varepsilon\sim \chi$ . At $\chi\gtrsim 1$  the energy of emitted photon is $ \omega\sim\varepsilon$.
This means that the noncommutativity of operators of the particle dynamical variables can be disregarded, while their commutators with the operators associated with the field of the radiated photons should be accounted for.
In this case operator formulation of quantum mechanics is particularly convenient. More details on the quasiclassical operator approach are given in books \cite{Baier1998,Landau_4}. By using this method  Baier and Katkov obtained following expression for the emission probability
\begin{equation}\label{PRB}
dw_{rad}=\frac{\alpha}{\left(2\pi\right)^{2}}\frac{d^{3}\mathbf{k}}{\omega}\int dt_{1}\int dt_{2}R_{2}^{*}R_{1}\exp\left[-i\frac{\varepsilon\left(kx_{2}-kx_{1}\right)}{\varepsilon'}\right],
\end{equation}
where $k^{\mu}=\omega\left\{ 1,\mathbf{n}\right\} $ and $x^{\mu}=\left\{ t,\mathbf{r}(t)\right\} $
are the 4-momentum and 4-coordinate of the emitted photon. The indices
1 and 2 denote the dependence on the radiation time moments $t_{1}$
and $t_{2}$ along $\mathbf{n}$ direction, respectively, $\mathbf{n}$ is
the radiation direction, $\varepsilon$ and $\varepsilon'$ the electron
energies before and after emission, respectively, and
\begin{align}
\label{R}
R(t) & =\varphi_{f}^{+}(\bm{\zeta}_f)\left[A(t)+i\bm{\sigma}\cdot \mathbf{B}(t)\right]\varphi_{i}(\bm{\zeta}_i),
\end{align}
where $\varphi_{i}$ and $\varphi_{f}$ are the two-component spinors
that describe the initial and final polarization states of the electron,
respectively. The unit vectors $\bm{\zeta}_i$ and $\bm{\zeta}_f$
are the corresponding polarization vectors. 
Taking into account Eq.(\ref{R}), 
we obtain
\begin{align}\label{R1R2} \nonumber
R_{2}^{*}R_{1} & =\frac{1}{4}\textrm{Tr}\left[\left(1+\bm{\zeta}_i\cdot\bm{\sigma}\right)\left(A_{2}^{*}-i\bm{\sigma}\cdot \mathbf{B}_2^{*}\right)\left(1+\bm{\zeta}_f\cdot \bm{\sigma}\right)\left(A_{1}+i\bm{\sigma}\cdot\mathbf{B}_1\right)\right]\\ \nonumber
 & =\frac{1}{2}\left[A_{1}A_{2}^{*}\left(1
 +\bm{\zeta}_i\cdot\bm{\zeta}_f\right)
 +\mathbf{B}_1\cdot\mathbf{B}_2^{*}\left(1
 -\bm{\zeta}_i\cdot\bm{\zeta}_f\right)\right.\\ \nonumber
 & +i\left(\bm{\zeta}_f-\bm{\zeta}_i\right)\cdot(\mathbf{B}_1\times\mathbf{B}_2^{*})+i\left(\bm{\zeta}_i+\bm{\zeta}_f\right)\cdot\left(\mathbf{B}_1A_{2}^{*}-A_{1}\mathbf{B}_2^{*}\right)\\  \nonumber
 & -\left(A_{1}\mathbf{B}_2^{*}+\mathbf{B}_1A_{2}^{*}\right)\cdot(\bm{\zeta}_i\times\bm{\zeta}_f)\\
 & \left.+\left(\bm{\zeta}_i\cdot\mathbf{B}_2^{*}\right)
 \left(\bm{\zeta}_f\cdot\mathbf{B}_1\right)
 +\left(\bm{\zeta}_i\cdot\mathbf{B}_1\right)
 \left(\mathbf{B}_2^{*}\cdot\bm{\zeta}_f\right)\right],
\end{align}
where the expressions of $A(t)$ and $\mathbf{B}(t)$ are
\begin{align}\label{AB}\nonumber
A(t) & =\frac{\mathbf{e}^{*}\cdot\mathbf{p}(t)}{2\sqrt{\varepsilon\varepsilon'}}\left[\left(\frac{\varepsilon'+m}{\varepsilon+m}\right)^{1/2}+\left(\frac{\varepsilon+m}{\varepsilon'+m}\right)^{1/2}\right],\\ \nonumber
\mathbf{B}(t)&=\frac{1}{2\sqrt{\varepsilon\varepsilon'}}\left[\left(\frac{\varepsilon'+m}{\varepsilon+m}\right)^{1/2}\mathbf{e}^{*}\times\mathbf{p}(t)+\left(\frac{\varepsilon+m}{\varepsilon'+m}\right)^{1/2}\mathbf{e}^{*}\times\right.\\
&\left.\left(\mathbf{p}(t)-\mathbf{k}\right)\right],
\end{align}
with $\mathbf{p}(t)=\gamma m \bm{\upsilon} $ being the momentum of the
electron, $\gamma=\varepsilon/m$ the Lorenz factor, $\mathbf{e}$
the polarization vector of the emitted photon. This expression can be used for calculation of any radiation characteristics,
including polarization and spin characteristics.

In LCFA the time of radiation in the given direction
 is much shorter than the time characteristic of particle motion,
and the variation of the external field acting on the particle at
the formation length can be neglected. In this case, it is convenient
to introduce the following variables
\begin{equation}\label{t1t2}
t=\left(t_{1}+t_{2}\right)/2,\tau=t_{2}-t_{1},
\end{equation}
and the functions in the probability expression expand over $\tau$: 
\begin{align}\nonumber \label{LCFA}
\mathbf{v}(t\pm\tau/2) & =\mathbf{v}(t)\pm\mathbf{w}\tau/2+\mathbf{\dot{w}}\tau^{2}/8+\cdots,\\
\mathbf{r}(t\pm\tau/2) & =\mathbf{r}(t)\pm\mathbf{v}\tau/2+\mathbf{w}\tau^{2}/8\pm\mathbf{\dot{w}}\tau^{3}/48+\cdots,
\end{align}
with $\mathbf{w}$ being the acceleration of electron.  Taking into account
that the produced particles are ultrarelativistic, one obtains  with an 
accuracy up to the terms $\sim O\left(1/\gamma^{2}\right)$
\begin{equation}
\mathbf{v\cdot w}=O\left(1/\gamma^{2}\right),\mathbf{n\cdot \dot{w}}=-w^{2}.
\end{equation}
Then
\begin{align}\nonumber\label{v1v2}
\mathbf{v_{1}v_{2}} & =1-\frac{1}{\gamma^{2}}-\frac{w^{2}\tau^{2}}{2},\\
kx_{2}-kx_{1} & =\omega\tau\left(1-\mathbf{n\cdot v}+w^{2}\tau^{2}/24\right).
\end{align}
For further calculation of probability $dw_{rad}$ in Eq. (\ref{PRB}), we introduce $\beta$, an angle
between the plane $\left(\mathbf{v},\mathbf{w}\right)$ and
vector $\mathbf{n}$; $\psi$, an angle between the projection
of vector $\mathbf{n}$ on the plane $\left(\mathbf{v},\mathbf{w}\right)$
and vector $\mathbf{v}$. The scalar combinations involving vector $\bf{n}$ have the form
\begin{align}\nonumber\label{n}
&\mathbf{n\cdot v}  =v\cos\beta\cos\psi,\\\nonumber
&\mathbf{n\cdot w_{\bot}}  =w_{\bot}\cos\beta\sin\psi,\\
&\mathbf{n\cdot}\left[\mathbf{v}\times\mathbf{w}_{\bot}\right]  =vw_{\bot}\sin\beta.
\end{align}
Since the ultrarelativistic particle radiates mainly forward into
a narrow cone, the angles $\beta$ and $\psi$ are of the order
of $1/\gamma$. With the adopted accuracy
\begin{align}\nonumber\label{n_angle}
& 1-\mathbf{n\cdot v} =\left(\beta^{2}+\psi^{2}+1/\gamma^{2}\right)/2,\\
& \mathbf{n}\cdot\mathbf{s}  =\psi, \mathbf{n\cdot}\left[\mathbf{v}\times\mathbf{s}\right]  =\beta.
\end{align}
where $\mathbf{s}=\mathbf{w}/|\mathbf{w}|$. Using Eqs.(\ref{n_angle}) and Eq.(\ref{v1v2}) in  Eq.(\ref{PRB}),
the photon radiation probability per unit time, $dW_{rad}\equiv dw_{rad}/dt$, reads
\begin{eqnarray}\label{PRB2}
dW_{rad}&=&\frac{\alpha\omega}{\left(2\pi\right)^{2}}d\omega\int_{-\infty}^\infty \int_{-\infty}^\infty d\beta d\psi \int_{-\infty}^\infty d\tau R_{2}^{*}R_{1}\\ \nonumber
&&\exp\left\{ -i\frac{\varepsilon}{\varepsilon'}\omega\left[\left(\frac{\beta^{2}}{2}+\frac{\psi^{2}}{2}+\frac{1}{2\gamma^{2}}\right)\tau+\frac{w^{2}\tau^{3}}{24}\right]\right\}.
\end{eqnarray}
Because of the rapid decreasing of functions at large angles and time, the integration limits
have been extended to infinity.

To investigate the radiation of a polarized photon by a polarized
electron in the constant field, we project the photon polarization
on the unit vectors 
\begin{eqnarray}
\mathbf{e}_1&=&\mathbf{s}-(\mathbf{n}\cdot\mathbf{s})\mathbf{n}, \nonumber\\
\mathbf{e}_2&=&\left[\mathbf{n}\times \mathbf{s}\right].
\end{eqnarray}
We shall proceed to the calculation
of Eq.(\ref{PRB2}) by integrating over $\tau$ and all angles.
Substituting the expressions in Eq.(\ref{R1R2}) into the radiation
probability and integrating over $\tau$ and all angles with the integrals shown in Appendix \ref{Appendix1},
we obtain the  polarization matrix of radiation probability per unit time:
\begin{align*}
dW_{11}+ dW_{22} & =\frac{C_0}{2}d\omega\left\{ \left[\frac{\varepsilon^{2}+\varepsilon'^{2}}{\varepsilon'\varepsilon}\textrm{K}_{\frac{2}{3}}\left(z_{q}\right)-\int_{z_q}^{\infty}dx\textrm{K}_{\frac{1}{3}}\left(x\right)\right]\right.\\
 & +\left[2\textrm{K}_{\frac{2}{3}}\left(z_{q}\right)-\int_{z_{q}}^{\infty}dx\textrm{K}_{\frac{1}{3}}\left(x\right)\right]\bm{\zeta}_i\cdot\bm{\zeta}_f\\
 & -\left[\frac{\omega}{\varepsilon}\bm{\zeta}_i\cdot\mathbf{b}+\frac{\omega}{\varepsilon'}\bm{\zeta}_f\cdot\mathbf{b}\right]\textrm{K}_{\frac{1}{3}}\left(z_{q}\right)\\
 & +\left.\frac{\omega^2}{\varepsilon'\varepsilon}\left[\textrm{K}_{\frac{2}{3}}\left(z_{q}\right)-\int_{z_{q}}^{\infty}dx\textrm{K}_{\frac{1}{3}}\left(x\right)\right]\left(\bm{\zeta}_i\cdot\mathbf{\hat{v}}\right)\left(\bm{\zeta}_f\cdot\mathbf{\hat{v}}\right)\right\},
\end{align*}
\begin{align*}
dW_{12}+dW_{21} & =\frac{C_0}{2}d\omega\left\{ \frac{\varepsilon^{2}-\varepsilon'^{2}}{2\varepsilon'\varepsilon}\textrm{K}_{\frac{2}{3}}\left(z_{q}\right)\left(\hat{\mathbf{v}}\left[\bm{\zeta}_f\times\bm{\zeta}_i\right]\right)\right.\\
 & +\left[\frac{\omega}{\varepsilon'}\left(\bm{\zeta}_i\cdot\mathbf{\mathbf{s}}\right)+\frac{\omega}{\varepsilon}\left(\bm{\zeta}_f\cdot\mathbf{s}\right)\right]\textrm{K}_{\frac{1}{3}}\left(z_{q}\right)\\
 & -\left.\frac{\omega^{2}}{2\varepsilon'\varepsilon}\int_{z_{q}}^{\infty}dx\textrm{K}_{\frac{1}{3}}\left(x\right)\left[\left(\bm{\zeta}_i\cdot\mathbf{s}\right)\left(\bm{\zeta}_f\cdot\mathbf{b}\right)+\left(\bm{\zeta}_i\cdot\mathbf{b}\right)\left(\bm{\zeta}_f\cdot\mathbf{s}\right)\right]\right\},
\end{align*}
\begin{align*}
dW_{12}-dW_{21} & =i\frac{C_0}{2}d\omega\left\{ \frac{\varepsilon^{2}-\varepsilon'^{2}}{2\varepsilon'\varepsilon}\textrm{K}_{\frac{1}{3}}\left(z_{q}\right)\left(\mathbf{s}\cdot\left[\bm{\zeta}_f\times\bm{\zeta}_i\right]\right)\right.\\
 & +\left(-\frac{\varepsilon^{2}-\varepsilon'^{2}}{\varepsilon'\varepsilon}\textrm{K}_{\frac{2}{3}}\left(z_{q}\right)+\frac{\omega}{\varepsilon}\int_{z_{q}}^{\infty}dx\textrm{K}_{\frac{1}{3}}\left(x\right)\right)\left(\bm{\zeta}_i\cdot\mathbf{\hat{v}}\right)\\
 & +\left(-\frac{\varepsilon^{2}-\varepsilon'^{2}}{\varepsilon'\varepsilon}\textrm{K}_{\frac{2}{3}}\left(z_{q}\right)+\frac{\omega}{\varepsilon'}\int_{z_{q}}^{\infty}dx\textrm{K}_{\frac{1}{3}}\left(x\right)\right)\left(\bm{\zeta}_f\cdot\mathbf{\hat{v}}\right)\\
 & +\left.\frac{\omega^{2}}{2\varepsilon'\varepsilon}\textrm{K}_{\frac{1}{3}}\left(z_{q}\right)\left[\left(\bm{\zeta}_i\cdot\mathbf{\hat{v}}\right)\left(\bm{\zeta}_f\cdot\mathbf{b}\right)+\left(\bm{\zeta}_i\cdot\mathbf{b}\right)\left(\bm{\zeta}_f\cdot\mathbf{\hat{v}}\right)\right]\right\},
\end{align*}
\begin{align}\nonumber \label{PRB_ELE}
dW_{11}-dW_{22} & =\frac{C_0}{2}d\omega\left\{ \textrm{K}_{\frac{2}{3}}\left(z_{q}\right)+\frac{\varepsilon^{2}+\varepsilon'^{2}}{2\varepsilon'\varepsilon}\textrm{K}_{\frac{2}{3}}\left(z_{q}\right)\bm{\zeta}_i\cdot\bm{\zeta}_f\right.\\\nonumber
 & -\left[\frac{\omega}{\varepsilon'}\left(\bm{\zeta}_i\cdot\mathbf{b}\right)+\frac{\omega}{\varepsilon}\left(\bm{\zeta}_f\cdot\mathbf{b}\right)\right]\textrm{K}_{\frac{1}{3}}\left(z_{q}\right)\\\nonumber
 & +\frac{\omega^{2}}{2\varepsilon'\varepsilon}\left(-\textrm{K}_{\frac{2}{3}}\left(z_{q}\right)\left(\bm{\zeta}_i\cdot\mathbf{\hat{v}}\right)\left(\bm{\zeta}_f\cdot\mathbf{\hat{v}}\right)\right.\\
 & +\left.\left.\int_{z_{q}}^{\infty}dx\textrm{K}_{\frac{1}{3}}\left(x\right)\left[\left(\bm{\zeta}_i\cdot\mathbf{b}\right)\left(\bm{\zeta}_f\cdot\mathbf{b}\right)-\left(\bm{\zeta}_i\cdot\mathbf{\mathbf{s}}\right)\left(\bm{\zeta}_f\cdot\mathbf{s}\right)\right]\right)\right\},
\end{align}
where $z_{q}=\frac{2}{3}\frac{\omega}{\chi\varepsilon'}$, $C_0=\frac{\alpha}{\sqrt{3}\pi\gamma^{2}}$ and $\mathbf{\hat{v}}=\mathbf{v}/\left|\mathbf{v}\right|$,
$\mathbf{b}=\mathbf{\hat{v}}\times\mathbf{s}$.
The radiation probability including all the polarization and spin characteristic takes the form
\begin{align}\label{PRB_tot_angle}
 dW_{rad}=\frac{1}{2}\left(F_0+\xi_1F_1+\xi_2F_2+\xi_3F_3\right),
\end{align}
where $F_0= dW_{11}+ dW_{22}$, $F_1= dW_{12}+ dW_{21}$, $F_2=i\left( dW_{12}- dW_{21}\right)$, $F_3= dW_{11}- dW_{22}$,
and the 3-vector $\bm{\xi}=\left(\xi_1,\xi_2,\xi_3\right)$ is the Stokes parameter of emitted photon defined with respect to $\mathbf{e}_1 $ and $\mathbf{e}_2$.
For an arbitrarily polarised photon  with polarisation vector $\mathbf{e}=a_{1}\mathbf{e}_1+a_{2}\mathbf{e}_2$  Stokes parameters are given by
\begin{align}
\xi_{1}=a_{1}a_{2}^{\ast}+a_{2}a_{1}^{\ast};\;\xi_{2}=i\left(a_{1}a_{2}^{\ast}-a_{2}a_{1}^{\ast}\right);\;\xi_{3}=\left|a_{1}\right|^{2}-\left|a_{2}\right|^{2}
\end{align}
After summing over the  polarization of emitted photon, we get
\begin{align*}
& dW_{rad}(\bm{\zeta}_i,\bm{\zeta}_f)  =a+\mathbf{b}\cdot\bm{\zeta}_f\\
&a =\frac{C_0}{2}d\omega\left\{ \frac{\varepsilon^{2}+\varepsilon'^{2}}{\varepsilon'\varepsilon}\textrm{K}_{\frac{2}{3}}\left(z_{q}\right)-\int_{z_q}^{\infty}dx\textrm{K}_{\frac{1}{3}}\left(x\right)-\frac{\omega}{\varepsilon}\bm{\zeta}_i\cdot\mathbf{b}\textrm{K}_{\frac{1}{3}}\left(z_{q}\right)\right\},\\
 &\mathbf{b}=\frac{C_0}{2}d\omega\left\{
 \left[2\textrm{K}_{\frac{2}{3}}\left(z_{q}\right)-\int_{z_{q}}^{\infty}dx\textrm{K}_{\frac{1}{3}}\left(x\right)\right]\bm{\zeta}_i-\frac{\omega}{\varepsilon'}\textrm{K}_{\frac{1}{3}}\left(z_{q}\right)\mathbf{b}\right.\\
 & +\left.\frac{\omega^2}{\varepsilon'\varepsilon}\left[\textrm{K}_{\frac{2}{3}}\left(z_{q}\right)-\int_{z_{q}}^{\infty}dx\textrm{K}_{\frac{1}{3}}\left(x\right)\right]\left(\bm{\zeta}_i\cdot\mathbf{\hat{v}}\right)\mathbf{\hat{v}}\right\},\\
\end{align*}
where $\bm{\zeta}_f$ is the final electron polarization defined by the detector.
The final
polarization vector of the electron resulting from the scattering process itself is
\begin{widetext}
\begin{equation}\label{zeta_f_R}
\bm{\zeta}_f^R=\frac{\mathbf{b}}{a}=\frac{
 \left[2\textrm{K}_{\frac{2}{3}}\left(z_{q}\right)-\int_{z_{q}}^{\infty}dx\textrm{K}_{\frac{1}{3}}\left(x\right)\right]\bm{\zeta}_i-\frac{\omega}{\varepsilon'}\textrm{K}_{\frac{1}{3}}\left(z_{q}\right)\mathbf{b}+\frac{\omega^2}{\varepsilon'\varepsilon}\left[\textrm{K}_{\frac{2}{3}}\left(z_{q}\right)-\int_{z_{q}}^{\infty}dx\textrm{K}_{\frac{1}{3}}\left(x\right)\right]\left(\bm{\zeta}_i\cdot\mathbf{\hat{v}}\right)\mathbf{\hat{v}}}{ \frac{\varepsilon^{2}+\varepsilon'^{2}}{\varepsilon'\varepsilon}\textrm{K}_{\frac{2}{3}}\left(z_{q}\right)-\int_{z_q}^{\infty}dx\textrm{K}_{\frac{1}{3}}\left(x\right)-\frac{\omega}{\varepsilon}\bm{\zeta}_i\cdot\mathbf{b}\textrm{K}_{\frac{1}{3}}\left(z_{q}\right)}.
\end{equation}
\end{widetext}
Taking the sum over the final electron polarizations, the radiation probability maintains the same form as Eq.(\ref{PRB_tot_angle}) but with the following coefficients:
\begin{align*}
\widetilde{F}_0 & =C_0d\omega\left\{ \frac{\varepsilon^{2}+\varepsilon'^{2}}{\varepsilon'\varepsilon}\textrm{K}_{\frac{2}{3}}\left(z_{q}\right)-\int_{z_q}^{\infty}dx\textrm{K}_{\frac{1}{3}}\left(x\right)-\frac{\omega}{\varepsilon}\bm{\zeta}_i\cdot\mathbf{b}\textrm{K}_{\frac{1}{3}}\left(z_{q}\right)\right\},
\end{align*}
\begin{align*}
\widetilde{F}_1 & =C_0d\omega \frac{\omega}{\varepsilon'}\left(\bm{\zeta}_i\cdot\mathbf{\mathbf{s}}\right)\textrm{K}_{\frac{1}{3}}\left(z_{q}\right),
\end{align*}
\begin{align*}
\widetilde{F}_2 & =-C_0d\omega
 \left(-\frac{\varepsilon^{2}-\varepsilon'^{2}}{\varepsilon'\varepsilon}\textrm{K}_{\frac{2}{3}}\left(z_{q}\right)+\frac{\omega}{\varepsilon}\int_{z_{q}}^{\infty}dx\textrm{K}_{\frac{1}{3}}\left(x\right)\right)\left(\bm{\zeta}_i\cdot\mathbf{\hat{v}}\right),
\end{align*}
\begin{align} \label{PRB_ELE2}
\widetilde{F}_3& =C_0d\omega\left\{ \textrm{K}_{\frac{2}{3}}\left(z_{q}\right)-\frac{\omega}{\varepsilon'}\left(\bm{\zeta}_i\cdot\mathbf{b}\right)\textrm{K}_{\frac{1}{3}}\left(z_{q}\right)\right\}.
\end{align}
\begin{figure*}
    \includegraphics[width=0.8\textwidth]{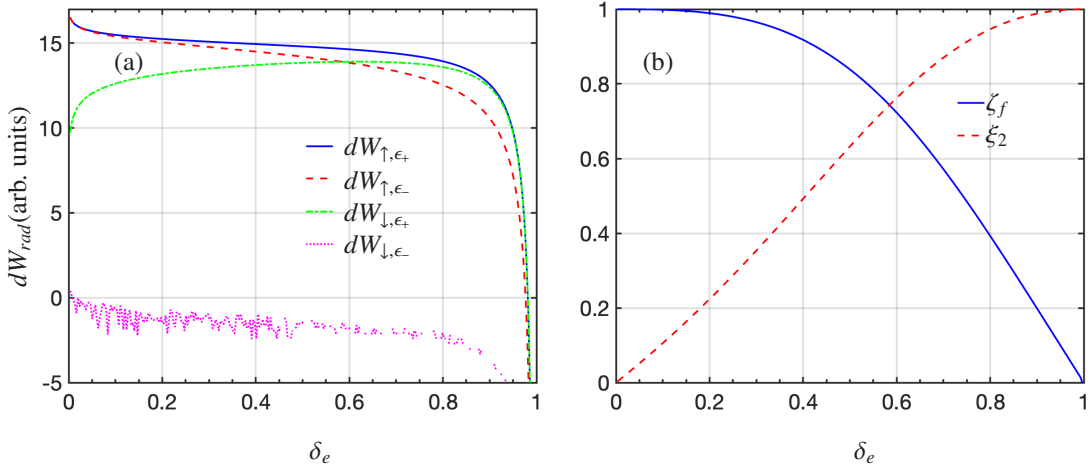}
    \begin{picture}(300,20)
	\put(-20,157){(a)}
	\put(180,157){(b)}
	\put(54,10){$\delta_e$}
	\put(260,10){$\delta_e$}
	\put(67,125){$dW_{\uparrow,\epsilon_+}$}
	\put(67,112){$dW_{\uparrow,\epsilon_-}$}
	\put(67,100){$dW_{\downarrow,\epsilon_+}$}
	\put(67,88){$dW_{\downarrow,\epsilon_-}$}
	\put(310,142){$\zeta_f$}
	\put(310,130){$\xi_2$}
	\put(-60,75){\rotatebox{90}{$dW_{rad}$(arb. units)}}
	\end{picture}
    \caption{(a) The radiation probability log$_{10}dW$ vs emitted photon energy $\delta_e=\omega_\gamma/\varepsilon_i$ for different final spins $\zeta_f\in\{\uparrow,\downarrow\}$ and photon polarizations $\epsilon\in\{\epsilon_+,\epsilon_-\}$, where $\epsilon_\pm=\frac{1}{\sqrt{2}}(\mathbf{e}_1\pm i\mathbf{e}_2)$. (b) The longitudinal polarization of electrons $\zeta_f=\frac{dW_\uparrow-dW_\downarrow}{dW_\uparrow+dW_\downarrow}$ (blue solid) and circular polarization of photons $\xi_2=\frac{dW_{\epsilon_+}-dW_{\epsilon_-}}{dW_{\epsilon_+}+dW_{\epsilon_-}}$ (red dashed) vs $\delta_e$; $\chi_e=1.0$, $\zeta_i=(0,0,1)$.}
    \label{Fig.rad}
\end{figure*}

The polarization of the emitted photon resulting from the scattering process itself takes the form $\xi_1^f=\widetilde{F}_1/\widetilde{F}_0$, $\xi_2^f=\widetilde{F}_2/\widetilde{F}_0$ and $\xi_3^f=\widetilde{F}_3/\widetilde{F}_0$. In linear Compton scattering the  polarization of photons is determined by the driving laser polarization, such that circularly polarized $\gamma$ photons can be obtained by linear Compton scattering of unpolarized electrons and a circularly polarized laser field. Otherwise in the nonlinear regime, see Eq.(\ref{PRB_ELE2}), the circular polarization of emitted photons is solely determined by
the longitudinal polarization of initial electrons $\xi_2\sim\bm{\zeta}_i\cdot \mathbf{e}_v$. Thus, circularly polarized $\gamma$-photons can be generated with nonlinear Compton scattering only if electrons are initially longitudinally polarized. As an example we calculate the emission probabilities for an initially polarized electron, see Fig. \ref{Fig.rad}. When the electron emits a low energy photon, the probabilities $dW_{\uparrow,\epsilon_+}$ and $dW_{\uparrow,\epsilon_-}$ dominate, leading to a small circular polarization of emitted photons. In the high energy region, $dW_{\uparrow,\epsilon_+}$ and $dW_{\downarrow,\epsilon_+}$ play leading roles, generating highly polarized gamma photons. In particular, when the emitted photon takes away nearly all the energy of the initial electron $\delta_e\sim 1$, $\xi_2\rightarrow 1$, i.e., the helicity of the electron is transferred to the emitted photon.

After averaging over initial electron polarizations, Eq.(\ref{PRB_ELE2}) becomes
\begin{align*}
\overline{F}_0 & =C_0d\omega\left\{ \frac{\varepsilon^{2}+\varepsilon'^{2}}{\varepsilon'\varepsilon}\textrm{K}_{\frac{2}{3}}\left(z_{q}\right)-\int_{z_q}^{\infty}dx\textrm{K}_{\frac{1}{3}}\left(x\right)\right\},
\end{align*}
\begin{align}\label{PRB_ELE3}
\overline{F}_1 & =0, \quad \overline{F}_2  =0, \quad\overline{F}_3 =C_0d\omega \textrm{K}_{\frac{2}{3}}\left(z_{q}\right),
\end{align}
which indicates that the emitted photon is always linearly polarized when electrons spin is unresolved.

\section{Spin and polarization resolved pair production probability}

We start from the general form of pair production probability given
in Ref. \cite{Baier1998}:
\begin{eqnarray}\label{PPP}
dW_{p}&=&\frac{dw_{p}}{dt}=\frac{\alpha}{\left(2\pi\right)^{2}}\frac{d^{3}p_{-}}{\omega}\int d\tau R_{p}^{*}\left(t-\frac{\tau}{2}\right)R_{p}\left(t+\frac{\tau}{2}\right)\nonumber\\
&\times&\exp\left\{ i\frac{\varepsilon}{\varepsilon_{f}}\left[kx\left(t+\frac{\tau}{2}\right)-kx\left(t-\frac{\tau}{2}\right)\right]\right\} ,
\end{eqnarray}
where $k^{\mu}=\omega\left(1,\mathbf{n}\right)$ and $x^{\mu}=\left(t,\mathbf{r}\left(t\right)\right)$
are four momentum and coordinate of the incoming photon, respectively, $\mathbf{n}$
is the unit vector in the photon propagation direction, which can be written
as $\mathbf{n}=\cos\psi\cos\beta\hat{\bm{\upsilon}} +\sin\psi\cos\beta\mathbf{s}+\sin\beta\mathbf{b}$
in the angle reference system $\left(\hat{\bm{\upsilon}},\mathbf{s},\mathbf{b}\right)$.
Here, $\hat{\bm{\upsilon}}$ is the unit vector along velocity of produced particles,
$\mathbf{s}$ the unit vector along transverse component
of acceleration $\mathbf{w}$, and $\mathbf{b}=\mathbf{\hat{\bm{\upsilon}}}\times\mathbf{s}$.
For ultrarelativistic particles, the angle between vector $\mathbf{n}$ and
$\bm{\upsilon} $ is of the order $1/\gamma$, therefore $\psi,\beta\sim1/\gamma$.
$\varepsilon_{+}$ and $\varepsilon_{-}$ are the energy of the created
positron and electron, respectively. The integration is performed
over the electron momentum $\mathbf{p}_{-}=\gamma m\bm{\upsilon}$. The expression
of $R_{p}\left(t\right)$ is represented with the form
\begin{equation}\label{RR}
R_{p}\left(t\right)=i\varphi_{s}^{+}\left(\bm{\zeta} (t)\right)\left(A\left(t\right)-i\bm{\sigma}\cdot\mathbf{B}\left(t\right)\right)\varphi_{\bar{s}}\left(\bm{\zeta}'\left(t\right)\right),
\end{equation}
$\varphi_{s}\left(\bm{\zeta}\right)$ and $\varphi_{\bar{s}}\left(\bm{\zeta}'\right)$
are the two-component spinors describing polarizations for particle
and antiparticle, with $\bm{\zeta}\left(\bm{\zeta}=\bm{\zeta}_{-}\right)$
and $\mathbf{\zeta}'\left(\mathbf{\zeta}'=-\bm{\zeta}_{+}\right)$ being the
corresponding polarization vectors.
With  Eq.(\ref{RR}), 
we obtain
\begin{align}\label{R1R22}\nonumber
R_{p}(t_{2})R_{p}^{*}(t_{1}) & =\frac{1}{2}\left[A_{1}^{*}A_{2}(1-\bm{\zeta}_{-}\cdot\bm{\zeta}_{+})+\mathbf{B_{1}^{*}}\mathbf{B}_2(1+\bm{\zeta}_{-}\cdot\bm{\zeta}_{+})\right.\\\nonumber
 & +(\bm{\zeta}_{-}\times\bm{\zeta}_{+})\cdot\left(\mathbf{B}_{1}^{*}A_{2}+\mathbf{B}_2A_{1}^{*}\right)\\\nonumber
 & -(\bm{\zeta}_{-}\cdot\mathbf{B_{1}^{*}})(\bm{\zeta}_{+}\cdot\mathbf{B}_2)-(\bm{\zeta}_{+}\cdot\mathbf{B_{1}^{*}})(\bm{\zeta}_{-}\cdot\mathbf{B}_2)\\\nonumber
 & +i(\bm{\zeta}_{-}-\bm{\zeta}_{+})\cdot\left(\mathbf{B_{1}^{*}}A_{2}-\mathbf{B}_2A_{1}^{*}\right)\\
 &\left.-i(\bm{\zeta}_{+}+\bm{\zeta}_{-})\cdot(\mathbf{B_{1}^{*}}\times\mathbf{B}_2)\right].
\end{align}
where
\begin{align*}
&A=N\mathbf{e}\cdot\left(\mathbf{k}\times\mathbf{p}\left(t\right)\right),\\
&\mathbf{B}\left(t\right)=N\left\{ \mathbf{e}\left[\left(\varepsilon'+m\right)\left(\varepsilon+m\right)-\mathbf{p}'\left(t\right)\cdot\mathbf{p}\left(t\right)\right]\right.\\
&\quad\quad\quad\left.+\left(\mathbf{e}\cdot\mathbf{p}\left(t\right)\right)\left(\mathbf{p}'\left(t\right)-\mathbf{p}\left(t\right)\right)\right\},\\
&N=\left[4\varepsilon\varepsilon'\left(\varepsilon+m\right)\left(\varepsilon'+m\right)\right]^{-1/2},
\end{align*}
with $\mathbf{e}$ being the photon polarization vector.

From now on we shall investigate pair production in a strong laser field
($a_{0}\gg1$), where the LCFA is
valid. In this case, 
the field inhomogeneity can be neglected when calculating the pair
production rate at time $t$. Using LCFA, the terms of $\bm{\upsilon}\left(t\pm\frac{\tau}{2}\right)$ and $\mathbf{r}\left(t\pm\frac{\tau}{2}\right)$ entering the pair production probability can be expanded as Eq. (\ref{LCFA}). Repeating the same steps for calculating the radiation probability, the pair production probability can be obtained with an accuracy of $\sim 1/\gamma$.
Specifically, projecting the photon
polarization on the unit vectors
$\bm{e}_1=\left(\bm{n}\times\bm{H}+\bm{E}_\perp\right)/\left(|\bm{n}\times\bm{H}+\bm{E}_\perp|\right)$ and $\bm{e}_2=\bm{n}\times\bm{e}_1$, 
substituting Eq.~(\ref{R1R22}) into the pair production probability Eq.~(\ref{PPP}), converting to angles $\psi$ and $\beta$ with Eq. (\ref{n}) and integrating over $\tau$ and the solid angle, see Appendix \ref{Appendix1}, one can obtain the electron spin and photon polarization resolved pair production probability for a photon with energy $\omega$ and Stokes parameters  $\xi_i (i=1,2,3)$ :
\begin{align}\nonumber
 dW_{p} & =\frac{1}{2}\left(dW_{11}+dW_{22}\right)+\frac{\xi_{1}}{2}\left(dW_{11}-dW_{22}\right)\\\nonumber
 & -i\frac{\xi_{2}}{2}\left(dW_{21}-dW_{12}\right)+\frac{\xi_{3}}{2}\left(dW_{11}-dW_{22}\right)\\
 & =\frac{1}{2}\left(G_{0}+\xi_{1}G_{1}+\xi_{2}G_{2}+\xi_{3}G_{3}\right),
\end{align}
where
\begin{align*}\nonumber
G_0&  =\frac{\overline{C}_{0}}{2}d\varepsilon\Bigg\{\left\{ \int_{z_{p}}^{\infty}dx\textrm{K}_{\frac{1}{3}}\left(x\right)+\frac{\varepsilon_{+}^{2}+\varepsilon^{2}}{\varepsilon_{+}\varepsilon}\textrm{K}_{\frac{2}{3}}\left(z_{p}\right)\right\} \\\nonumber
 & +\left\{ \int_{z_{p}}^{\infty}dx\textrm{K}_{\frac{1}{3}}\left(x\right)-2\textrm{K}_{\frac{2}{3}}\left(z_{p}\right)\right\} \left(\bm{\zeta}_{-}\cdot\bm{\zeta}_{+}\right)\\\nonumber
 & +\left[\frac{\omega}{\varepsilon_{+}}\left(\bm{\zeta}_{+}\cdot\mathbf{b}\right)-\frac{\omega}{\varepsilon}\left(\bm{\zeta}_{-}\cdot\mathbf{b}\right)\right]\textrm{K}_{\frac{1}{3}}\left(z_{p}\right)\\
 & +\left\{ \frac{\varepsilon_{+}^{2}+\varepsilon^{2}}{\varepsilon\varepsilon_{+}}\int_{z_{p}}^{\infty}dx\textrm{K}_{\frac{1}{3}}\left(x\right)-\frac{\left(\varepsilon_{+}-\varepsilon\right)^{2}}{\varepsilon\varepsilon_{+}}\textrm{K}_{\frac{2}{3}}\left(z_{p}\right)\right\} \\
&\quad\left(\bm{\zeta}_{-}\cdot \hat{\bm{\upsilon}}\right)\left(\bm{\zeta}_{+}\cdot \hat{\bm{\upsilon}}\right)\Bigg\}
\end{align*}
\begin{align*}\nonumber
G_3 & =\frac{\overline{C}_{0}}{2}d\varepsilon\Bigg\{-\textrm{K}_{\frac{2}{3}}\left(z_{p}\right)+\frac{\varepsilon_{+}^{2}+\varepsilon^{2}}{2\varepsilon_{+}\varepsilon}\textrm{K}_{\frac{2}{3}}\left(z_{p}\right)\left(\bm{\zeta}_{-}\cdot\bm{\zeta}_{+}\right)\\\nonumber
 & +\left[-\frac{\omega}{\varepsilon}\left(\bm{\zeta}_{+}\cdot\mathbf{b}\right)+\frac{\omega}{\varepsilon_{+}}\left(\bm{\zeta}_{-}\cdot\mathbf{b}\right)\right]\textrm{K}_{\frac{1}{3}}\left(z_{p}\right)\\\nonumber
 & -\frac{\left(\varepsilon_{+}-\varepsilon\right)^{2}}{2\varepsilon_{+}\varepsilon}\textrm{K}_{\frac{2}{3}}\left(z_{p}\right)\left(\bm{\zeta}_{-}\cdot\hat{\bm{\upsilon}}\right)\left(\bm{\zeta}_{+}\cdot\hat{\bm{\upsilon}}\right)\\
 & +\frac{\omega^{2}}{2\varepsilon_{+}\varepsilon}\int_{z_{p}}^{\infty}dx\textrm{K}_{\frac{1}{3}}\left(x\right)\left[\left(\bm{\zeta}_{-}\cdot\mathbf{b}\right)\left(\bm{\zeta}_{+}\cdot\mathbf{b}\right)-\left(\bm{\zeta}_{-}\cdot\mathbf{s}\right)\left(\bm{\zeta}_{+}\cdot\mathbf{s}\right)\right]\Bigg\}
\end{align*}
\begin{align*}\nonumber
G_1 & =\frac{\overline{C}_{0}}{2}d\varepsilon\Bigg\{-\frac{\varepsilon_{+}^{2}-\varepsilon^{2}}{2\varepsilon_{+}\varepsilon}\textrm{K}_{\frac{2}{3}}\left(z_{p}\right)\hat{\bm{\upsilon}}\cdot(\bm{\zeta}_{+}\times\bm{\zeta}_{-})\\\nonumber
 & +\left[\frac{\omega}{\varepsilon}\left(\bm{\zeta}_{+}\cdot\mathbf{s}\right)-\frac{\omega}{\varepsilon_{+}}\left(\bm{\zeta}_{-}\cdot\mathbf{s}\right)\right]\textrm{K}_{\frac{1}{3}}\left(z_{p}\right)\\
 & -\frac{\omega^{2}}{2\varepsilon_{+}\varepsilon}\int_{z_{p}}^{\infty}dx\textrm{K}_{\frac{1}{3}}\left(x\right)\left\{ \left(\bm{\zeta}_{-}\cdot\mathbf{b}\right)\left(\bm{\zeta}_{+}\cdot\mathbf{s}\right)+\left(\bm{\zeta}_{-}\cdot\mathbf{s}\right)\left(\bm{\zeta}_{+}\cdot\mathbf{b}\right)\right\} \Bigg\}
\end{align*}
\begin{align}\nonumber
G_2 & =\frac{\overline{C}_{0}}{2}d\varepsilon\Bigg\{-\frac{\omega^{2}}{2\varepsilon_{+}\varepsilon}\textrm{K}_{\frac{1}{3}}\left(z_{p}\right)\left[\mathbf{s}\cdot(\bm{\zeta}_{-}\times\bm{\zeta}_{+})\right]\\\nonumber
 & +\left(\frac{\omega}{\varepsilon_{+}}\int_{z_{p}}^{\infty}dx\textrm{K}_{\frac{1}{3}}\left(x\right)+\frac{\varepsilon_{+}^{2}-\varepsilon^{2}}{\varepsilon_{+}\varepsilon}\textrm{K}_{\frac{2}{3}}\left(z_{p}\right)\right)\left(\bm{\zeta}_{+}\cdot\hat{\bm{\upsilon}}\right)\\\nonumber
 & +\left(\frac{\omega}{\varepsilon}\int_{z_{p}}^{\infty}dx\textrm{K}_{\frac{1}{3}}\left(x\right)-\frac{\varepsilon_{+}^{2}-\varepsilon^{2}}{\varepsilon_{+}\varepsilon}\textrm{K}_{\frac{2}{3}}\left(z_{p}\right)\right)\left(\bm{\zeta}_{-}\cdot\hat{\bm{\upsilon}}\right)\\
 & -\frac{\varepsilon_{+}^{2}-\varepsilon^{2}}{2\varepsilon_{+}\varepsilon}\textrm{K}_{\frac{1}{3}}\left(z_{p}\right)\left[\left(\bm{\zeta}_{-}\cdot\hat{\bm{\upsilon}} \right)\left(\bm{\zeta}_{+}\cdot\mathbf{b}\right)+\left(\bm{\zeta}_{-}\cdot\mathbf{b}\right)\left(\bm{\zeta}_{+}\cdot\hat{\bm{\upsilon}} \right)\right]\Bigg\}.
\end{align}
\begin{figure*}
    \includegraphics[width=0.8\textwidth]{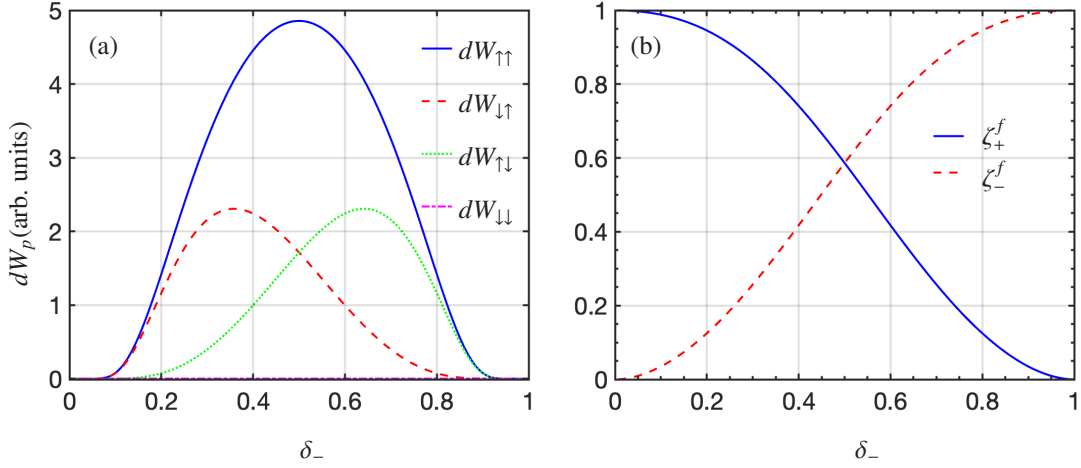}
    \begin{picture}(300,20)
	\put(-30,163){(a)}
	\put(175,163){(b)}
	\put(50,10){$\delta_-$}
	\put(260,10){$\delta_-$}
	\put(110,161){$dW_{\uparrow\uparrow}$}
	\put(110,141){$dW_{\downarrow\uparrow}$}
	\put(110,121){$dW_{\uparrow\downarrow}$}
	\put(110,101){$dW_{\downarrow\downarrow}$}
	\put(308,129){$\zeta^f_+$}
	\put(308,113){$\zeta^f_-$}
	\put(-60,75){\rotatebox{90}{$dW_p$(arb. units)}}
	\end{picture}
    \caption{(a) The pair production probability $dW_{\zeta_-\zeta_+}$ vs  relative energy of electron $\delta_-=\varepsilon/\omega$ for different final spins of electron $\zeta^f_-\in\{\uparrow,\downarrow\}$ and positron $\zeta^f_+\in\{\uparrow,\downarrow\}$. (b) The longitudinal polarization of electrons $\zeta^f_-=\frac{\sum_{i} dW_{\uparrow,i}-\sum_{i} dW_{\downarrow,i}}{\sum_{i}dW_{\uparrow,i}+\sum_{i}dW_{\downarrow,i}}$ (blue solid) and positrons $\zeta^f_+=\frac{\sum_{i} dW_{i,\uparrow}-\sum_{i} dW_{i,\downarrow}}{\sum_{i}dW_{i,\uparrow}+\sum_{i}dW_{i,\downarrow}}$ (red dashed) vs $\delta_-$; $\chi_\gamma=1.0$, $\xi=(0,1,0)$.}
    \label{Fig.pair}
\end{figure*}
Here $\overline{C}_0=\frac{\alpha m^{2}d\varepsilon}{\sqrt{3}\pi\omega^{2}}$ and $z_p=\frac{2}{3\chi_\gamma}\frac{\omega^2}{\varepsilon_+\varepsilon_-}$. After taking the sum over positron polarizations, we arrive at the  results given in Ref. \cite{Baier1998}:
\begin{align*}\nonumber
\widetilde{G}_0&  =\overline{C}_0d\varepsilon\Bigg\{ \int_{z_{p}}^{\infty}dx\textrm{K}_{\frac{1}{3}}\left(x\right)+\frac{\varepsilon_{+}^{2}+\varepsilon^{2}}{\varepsilon_{+}\varepsilon}\textrm{K}_{\frac{2}{3}}\left(z_{p}\right)
 -\frac{\omega}{\varepsilon}\left(\bm{\zeta}_{-}\cdot\mathbf{b}\right)\textrm{K}_{\frac{1}{3}}\left(z_{p}\right)\Bigg\}
\end{align*}
\begin{align*}\nonumber
\widetilde{G}_3 & =\overline{C}_0d\varepsilon\Bigg\{-\textrm{K}_{\frac{2}{3}}\left(z_{p}\right)
 +\frac{\omega}{\varepsilon_{+}}\left(\bm{\zeta}_{-}\cdot\mathbf{b}\right)\textrm{K}_{\frac{1}{3}}\left(z_{p}\right)\Bigg\}
\end{align*}
\begin{align*}\nonumber
\widetilde{G}_1 & =-\overline{C}_0d\varepsilon\frac{\omega}{\varepsilon_{+}}\left(\bm{\zeta}_{-}\cdot\mathbf{s}\right)\textrm{K}_{\frac{1}{3}}\left(z_{p}\right)
\end{align*}
\begin{align}\nonumber
\widetilde{G}_2 & =\overline{C}_0d\varepsilon\Bigg\{ \left(\frac{\omega}{\varepsilon}\int_{z_{p}}^{\infty}dx\textrm{K}_{\frac{1}{3}}\left(x\right)-\frac{\varepsilon_{+}^{2}-\varepsilon^{2}}{\varepsilon_{+}\varepsilon}\textrm{K}_{\frac{2}{3}}\left(z_{p}\right)\right)\left(\bm{\zeta}_{-}\cdot\hat{\bm{\upsilon}} \right)\Bigg\}.
\end{align}
The polarization of the final electron can be expressed as
\begin{widetext}
\begin{equation}\label{PPE}
\bm{\zeta}_{-}^{f}=-\frac{\xi_{1}\textrm{K}_{\frac{1}{3}}\left(z_{p}\right)\frac{\omega}{\varepsilon_{+}}\mathbf{s}+\xi_{2}\hat{\bm{\upsilon}} \left(-\frac{\omega}{\varepsilon}\int_{z_{p}}^{\infty}dx\textrm{K}_{\frac{1}{3}}\left(x\right)+\frac{\varepsilon_{+}^{2}-\varepsilon^{2}}{\varepsilon\varepsilon_{+}}\textrm{K}_{\frac{2}{3}}\left(z_{p}\right)\right)+\left(\frac{\omega}{\varepsilon}-\xi_{3}\frac{\omega}{\varepsilon_{+}}\right)\mathbf{b}\textrm{K}_{\frac{1}{3}}\left(z_{p}\right)}{\int_{z_{p}}^{\infty}dx\textrm{K}_{\frac{1}{3}}\left(x\right)+\frac{\varepsilon^{2}+\varepsilon_{+}^{2}}{\varepsilon\varepsilon_{+}}\textrm{K}_{\frac{2}{3}}\left(z_{p}\right)-\xi_{3}\textrm{K}_{\frac{2}{3}}\left(z_{p}\right)}.
\end{equation}
\end{widetext}
In nonlinear Breit-Wheeler process, see Eq.(\ref{PPE}), the longitudinal polarization of the created electrons is solely determined by circular polarization of initial gamma photons $\zeta_\parallel\equiv\bm{\zeta}_-\cdot \mathbf{e}_v\propto \xi_2$. Thus, longitudinal polarized electrons can be generated via the nonlinear Breit-Wheeler process only if initial gamma photons are circularly polarized. As an example we calculate the pair production probabilities for an circularly polarized gamma photon, see Fig. \ref{Fig.pair}. When a low energy electron is created, the probabilities $dW_{\uparrow\uparrow}$ and $dW_{\downarrow\uparrow}$ dominate, leading to a small longitudinal polarization of created electrons. In the high energy region, $dW_{\uparrow\uparrow}$ and $dW_{\uparrow\downarrow}$ play leading roles, generating highly longitudinally polarized electron. In particular,  $\bm{\zeta}_{-}^{f}=\xi_2\mathbf{e}_v$ when $\varepsilon\approx\omega$. This is the case when helicity transfer occurs. For this reason, it is possible to produce longitudinally polarized positrons via the Breit-Wheeler process \cite{li2020production}.
After taking the sum over positron and electron polarizations, we get the spin unresolved pair production probability:
\begin{align*}\nonumber
\overline{G}_0&  =2\overline{C}_0d\varepsilon\Bigg\{ \int_{z_{p}}^{\infty}dx\textrm{K}_{\frac{1}{3}}\left(x\right)+\frac{\varepsilon_{+}^{2}+\varepsilon^{2}}{\varepsilon_{+}\varepsilon}\textrm{K}_{\frac{2}{3}}\left(z_{p}\right)\Bigg\},
\end{align*}
\begin{equation}
\overline{G}_3 =-2\overline{C}_0d\varepsilon\textrm{K}_{\frac{2}{3}}\left(z_{p}\right), \overline{G}_1=0,\overline{G}_2=0.
\end{equation}
As expected, the dependence on circular polarization of the photon is vanishing if the spin is unresolved.

\section{Conclusion}

Some of the results are already applied to the simulation codes in our previous works but without derivation \cite{Liyf2019,Chen2019,Liyf2020,li2020production,Wanf2020,Wan2020}. In this paper, we give a rigorous derivation of
the electron spin and photon polarization fully resolved radiation and pair production probabilities. Using Baier-Katkov semiclassical method, we obtain the complete set of the radiation  and pair production probabilities for arbitrary initial and final electron-positron spins and arbitrary polarization of the incoming and outgoing photons within the LCFA  and quasiclasscial approximation. The fully polarization-resolved  probabilities are written in a compact form, and are applicable for  arbitrary pulse shapes and polarization as long as $a_0\gg1$ and $\gamma\gg 1$.
Therefore, the formulas can be easily implemented into QED Monte Carlo simulation codes for investigating polarization effects during nonlinear Compton scattering and Breit-Wheeler processes, and are  essential for polarization effect studies in strong field QED precesses.

\section*{Acknowledgement}
This work has been supported by the National Natural Science Foundation of China (Grants No. 12074262) and the Program for Professor of Special Appointment (Eastern Scholar) at Shanghai Institutions of Higher Learning and Shanghai Rising-Star Program.

\appendix*

\section{Calculation of the integrals}\label{Appendix1}
The integration over $\tau$ in Eq.~(11) employs the following results:
\begin{align*}
\int_{-\infty}^{\infty}dx\cos\left(bx+ax^{3}\right)&=\frac{2}{3}\sqrt{\frac{b}{a}}\textrm{K}_{\frac{1}{3}}\left(\frac{2}{3\sqrt{3}}\frac{b^{\frac{3}{2}}}{\sqrt{a}}\right),\\
\int_{-\infty}^{\infty}dx\,x\,\sin\left(bx+ax^{3}\right)&=\frac{2}{3\sqrt{3}}\frac{b}{a}\textrm{K}_{\frac{2}{3}}\left(\frac{2}{3\sqrt{3}}\frac{b^{\frac{3}{2}}}{\sqrt{a}}\right),\\
\int_{-\infty}^{\infty}dx\,x^{2}\cos\left(bx+ax^{3}\right)&=-\frac{2}{9}\left(\frac{b}{a}\right)^{\frac{3}{2}}\textrm{K}_{\frac{1}{3}}\left(\frac{2}{3\sqrt{3}}\frac{b^{\frac{3}{2}}}{\sqrt{a}}\right).\numberthis
\end{align*}
Therefore, we have
\begin{align}\nonumber
I_{0}&=\int_{-\infty}^{\infty}d\tau\exp\left\{ -i\frac{\varepsilon\omega}{\varepsilon_{f}}\left[\left(1-\mathbf{n}\mathbf{v}\right)\tau+\frac{\tau^{3}}{24}\mathbf{w}^{2}\right]\right\}  \\\nonumber
&=\frac{2\sqrt{24}}{3w}\sqrt{\left(1-\mathbf{n}\mathbf{v}\right)}\textrm{K}_{\frac{1}{3}}\left(\xi\right),\\\nonumber
 & I_{1}=\int_{-\infty}^{\infty}d\tau\tau\exp\left\{ -i\frac{\varepsilon\omega}{\varepsilon_{f}}\left[\left(1-\mathbf{n}\mathbf{v}\right)\tau+\frac{\tau^{3}}{24}\mathbf{w}^{2}\right]\right\} \\\nonumber
 &=-i\frac{16}{\sqrt{3}}\frac{1}{w^{2}}\left(1-\mathbf{n}\mathbf{v}\right)\textrm{K}_{\frac{2}{3}}\left(\xi\right),\\\nonumber
 & I_{2}=\int_{-\infty}^{\infty}d\tau\tau^{2}\exp\left\{ -i\frac{\varepsilon\omega}{\varepsilon_{f}}\left[\left(1-\mathbf{n}\mathbf{v}\right)\tau+\frac{\tau^{3}}{24}\mathbf{w}^{2}\right]\right\} \\
 &=-\frac{2\left(24\right)^{\frac{3}{2}}}{9w^{3}}\left(1-\mathbf{n}\mathbf{v}\right)^{\frac{3}{2}}\textrm{K}_{\frac{1}{3}}\left(\xi\right).\numberthis
\end{align}
Integrating Eq.(14) over solid angles,  one obtains the integrals for deriving the radiation probability:
\begin{align*}
 \int d\Omega I_{0}& =\frac{4\sqrt{3}}{3}\pi\frac{\varepsilon'}{\varepsilon\omega}\int_{z_q}^{\infty}dx\textrm{K}_{\frac{1}{3}}\left(x\right),\\
 \end{align*}
 \begin{align*}
 \int d\Omega I_{1}& =-i\frac{8}{\sqrt{3}}\frac{1}{w}\pi\frac{1}{\gamma}\frac{\varepsilon'}{\varepsilon\omega}\textrm{K}_{\frac{1}{3}}\left(z_q\right),\\
\end{align*}
 \begin{align*}
\int d\Omega I_{2}& = -\frac{16}{\sqrt{3}w^{2}}\pi\frac{1}{\gamma^{2}}\frac{\varepsilon'}{\varepsilon\omega}\textrm{K}_{\frac{2}{3}}\left(z_q\right).\numberthis
\end{align*}
Similarly, the pair production probability can be obtained with the integration over $\tau$:
\begin{align}\nonumber
I_{0}&=\int_{-\infty}^{\infty}d\tau\exp\left\{ i\frac{\varepsilon\omega}{\varepsilon_{+}}\left[\left(1-\mathbf{n}\mathbf{v}\right)\tau+\frac{\tau^{3}}{24}\mathbf{w}^{2}\right]\right\}  \\\nonumber
&=\frac{2\sqrt{24}}{3w}\sqrt{\left(1-\mathbf{n}\mathbf{v}\right)}\textrm{K}_{\frac{1}{3}}\left(\xi\right),\\\nonumber
 \end{align}
 \begin{align}\nonumber
 & I_{1}=\int_{-\infty}^{\infty}d\tau\tau\exp\left\{ i\frac{\varepsilon\omega}{\varepsilon_{+}}\left[\left(1-\mathbf{n}\mathbf{v}\right)\tau+\frac{\tau^{3}}{24}\mathbf{w}^{2}\right]\right\} \\\nonumber
  &=i\frac{16}{\sqrt{3}}\frac{1}{w^{2}}\left(1-\mathbf{n}\mathbf{v}\right)\textrm{K}_{\frac{2}{3}}\left(\xi\right),\\\nonumber
 & I_{2}=\int_{-\infty}^{\infty}d\tau\tau^{2}\exp\left\{ i\frac{\varepsilon\omega}{\varepsilon_{+}}\left[\left(1-\mathbf{n}\mathbf{v}\right)\tau+\frac{\tau^{3}}{24}\mathbf{w}^{2}\right]\right\} \\
 &=-\frac{2\left(24\right)^{\frac{3}{2}}}{9w^{3}}\left(1-\mathbf{n}\mathbf{v}\right)^{\frac{3}{2}}\textrm{K}_{\frac{1}{3}}\left(\xi\right).
\end{align}
and the solid angle integrations give
\begin{align*}
 \int d\Omega I_{0}& =\frac{4\sqrt{3}}{3}\pi\frac{\varepsilon_+}{\varepsilon\omega}\int_{z_p}^{\infty}dx\textrm{K}_{\frac{1}{3}}\left(x\right),\\
 \int d\Omega I_{1}& =i\frac{8}{\sqrt{3}}\frac{1}{w}\pi\frac{1}{\gamma}\frac{\varepsilon_+}{\varepsilon\omega}\textrm{K}_{\frac{1}{3}}\left(z_p\right),\\
\int d\Omega I_{2}& = -\frac{16}{\sqrt{3}w^{2}}\pi\frac{1}{\gamma^{2}}\frac{\varepsilon_+}{\varepsilon\omega}\textrm{K}_{\frac{2}{3}}\left(z_p\right).
\end{align*}

\bibliography{RPP}
\end{document}